\RequirePackage{fix-cm}
\documentclass[smallextended,10pt]{svjour3}       
\smartqed  
\usepackage{graphicx}
\usepackage{multirow}
\usepackage{tabularx}
\newcolumntype{C}[1]{>{\centering\arraybackslash}m{#1}}
\newcolumntype{R}[1]{>{\RaggedLeft\arraybackslash}p{#1}}
\usepackage{hyperref}
\usepackage{amsmath,amssymb,amsfonts}
\usepackage{makecell}
\usepackage{subfig}
\usepackage{setspace}
\usepackage{float}
\usepackage{array,booktabs,ragged2e}

\usepackage{comment}
\usepackage{xcolor}
\usepackage{multirow}

\usepackage[normalem]{ulem} 

\begin{document}

\title{Analyzing Cooperative Monitoring and Dissemination of Critical Mobile Events by VANETs
}

\titlerunning{Analyzing Cooperative Monitoring and Dissemination of Critical Mobile Events}        

\author{Everaldo Andrade \and Aldri Santos \and Paulo D. Maciel Jr. \and Fernando Matos}


\institute{Everaldo Andrade \and Fernando Matos \at
                Dept. of Computer Science - Federal University of Para\'iba, Brazil \\
                \email{everaldo.andrade@ppgi.ci.ufpb.br; fernando@ci.ufpb.br}           
            \and
                Aldri Santos \at
                NR2 - Federal University of Paran\'a, Brazil\\
                \email{aldri@inf.ufpr.br}
            \and
                Paulo D. Maciel Jr. \at 
                Academic Unit of Informatics - Federal Institute of Para\'iba, Brazil \\
                \email{paulo.maciel@ifpb.edu.br}
}

\date{Received: date / Accepted: date}

\maketitle

\begin{abstract}
The treatment of mobile and simultaneous critical urban events requires effective actions by the appropriate authorities. Additionally it implies communication challenges in the speed and accuracy of their occurrence by the entities, as well as dealing with the dynamics and speed in these environments. Cooperative solutions with shared resources that address these challenges become a real option in helping to handle these events. This paper presents an evaluation of dynamic monitoring and collaborative dissemination supported by vehicular groups. It aims to analyze the impact of multiple mobile and fixed events in an urban environment on information propagation, considering barriers imposed by the events and the environment.
Differently from other studies in the literature, this work takes into account both fixed and mobile events, as well as simultaneous events.
NS3 results show that the evaluated system monitored at least 87\% and 51.5\% of the time for mobile and fixed events respectively, and delivered information over 77\% and 50\% of the time for those events, with average delay remains close to 0.3s in most scenarios. 
The results also reveal that a more continuous monitoring of the mobile events is highly dependent on the orientation of the vehicles. The main contribution of this work consists of the performance analysis of both fixed and mobile simultaneous events to support studies on how moving events impact on the dissemination and delivery of real-time data, and thus encouraging the development of new data dissemination protocols for VANETs.
\keywords{VANETs \and Mobile crowdsensing \and Dissemination \and Critical mobile events \and Mobility impact analysis}.
\end{abstract}

\section{Introduction}
\label{sec:intro}
The surging number of distinct critical events in urban areas, such as accidents, natural disasters, and civil unrest, has become a matter of concern to both the population and public authorities. These events compromise the operation of cities and can impact the citizens' daily life~\cite{Zhang2016} \cite{Costa2020}. Thus, researchers have developed emergency planning models to prevent critical events and/or mitigate their consequences in diverse scenarios
\cite{TangY2019} \cite{Cantillo2019} \cite{TangP2019} \cite{Amiri2020} \cite{Ferranti2019}. However, a variety of
events take place randomly over time and space, which hinders their treatment. In addition, an efficient handling of events requires an automatic and precise perception of its occurrence and immediate announcement to the authorities, so that they can take appropriate actions \cite{Immich2019}~\cite{Hilal2020} \cite{Chou2017}. Since the perception and the announcement of events mostly depend on human intervention, their efficient treatment is still a challenging task.

In pervasive environments, the collaborative potential between computational systems over urban hybrid infrastructures in Smart Cities enables the development and implementation of resilient solutions capable of handling critical urban events \cite{Shah2019}. Specifically, studies in the field of Vehicular Ad Hoc Networks (VANETs), has been carried out to avoid or mitigate events and their consequences in urban environments \cite{Astarita2020} \cite{Boukerche2019}.
VANETs facilitate the dissemination of data through Vehicle-to-Vehicle (V2V) and Vehicle-to-Infrastructure (V2I) communications \cite{Shrestha2018} \cite{Li2019}, supported by communication technologies such as IEEE 802.11p and LTE, and not facing resource limitations that are inherent in traditional mobile ad hoc networks (e.g. CPU, memory) \cite{khakpour_using_2017}. Moreover, the ever-increasing manufactured embedded cameras in vehicles and their pervasiveness in traffic lanes, turn VANETs into a natural crowdsensing environment useful for event detection and monitoring \cite{Kim2019}.

However, to achieve an efficient treatment of events, one must take into account the characteristics of both the urban event and the vehicular environment. As for the urban event, solutions must deal with: \textit{a) Mobility} \cite{Adam2020} -- the event can vary~its position over time; and \textit{b) Simultaneity} \cite{Palmieri2016} -- events occur simultaneously at different locations in the environment. In addition, the vehicular environment also imposes its own challenges: \textit{i) Traffic model} \cite{Yelure2020} \cite{Harri2009} -- the environment may hold~different traffic patterns depending on the time and the traffic regulations; \textit{ii) Node speed} \cite{Medeiros2019} -- high variability in vehicle speed directly~affects the communication; and \textit{iii) Density} \cite{Goyal2019} -- traffic density varies~over time and space due to the highly dynamic nature of~VANETs.

A couple of works in the literature have highlighted the importance of assessing the performance of critical data dissemination in VANETs for different goals, e.g. assistance in the treatment of traffic accidents \cite{Chou2017}, tracking of escaping vehicles \cite{khakpour_using_2017} or broadcasting emergency messages \cite{Ramakrishnan2017}. The study in \cite{Balakumar2019} analyzes the efficiency of dissemination techniques in meeting the environment requirements. In~\cite{Dar2010}, a performance analysis of data dissemination protocols applied to VANETs is carried out, where events are place at different locations in the test scenario. The impact of using different radio propagation models to disseminate emergency messages is investigated in \cite{Suthaputchakun2015}. In \cite{Paranjothi2020} a performance analysis to disseminate safety and non-safety messages in VANETs based on the fog paradigm is presented. However, these works do not consider important contexts of the events, such as mobility and multiple occurrences, do not consider hybrid communication in VANETs (V2V and V2I) or lack deeper analysis on the impact of the solution overhead in the network.

This work presents a study that aims to assess the impact of the mobility and simultaneity of urban events on the monitoring of critical events and the dissemination of real-time data in VANETs. 
The results of the study are expected to provide useful information to encourage and foster the development of new data dissemination protocols that deal with moving and simultaneous events in VANETs.
We use MINUET~\cite{Andrade2020}, a monitoring and dissemination system that operates in hybrid network (V2V and V2I), in which vehicles cooperate to monitor urban events and deliver their data to the competent authorities. Simulated experiments showed that the mobility aspect strongly influences on the monitoring, dissemination, and delivery of urban events' data. The monitoring of mobile events was perceived in 88\% of the simulated time against 56\% for fixed events, in scenarios of high traffic density; and respectively 87\% and 51.5\% in low traffic density scenarios. The dissemination and delivery of information were also impacted by the mobility of events; i.e. the MINUET was able to deliver data during 77\% of the time for mobile ones, against 50\% of the time for fixed ones.
Moreover, the orientation of the vehicles highly affects a more continuous monitoring of mobile events.
In contrast to other works in the literature, such as \cite{benkerdagh_cluster_2019}, \cite{siddiqua_icafe_2019} and \cite{Chou2017}, our work also focuses on mobile events. In addition, we also assess the performance of data dissemination of simultaneous events, differently from \cite{khakpour_using_2017} and \cite{Derder2019}, that considers them isolated. Therefore, the main contributions of this article are twofold:

\begin{enumerate}
    \item Performance study of both fixed and mobile simultaneous events in VANETs and comparison assessment regarding different clustering techniques to monitor events;
    \item Simulations carried out taking into account real urban traffic and hybrid communication (V2V and V2I).
\end{enumerate}


The remainder of this paper is organized as follows. Section \ref{sec:related} reviews closely related works. Section \ref{sec:minuet} describes the MINUET system and provides a model characterization. Section \ref{sec:setup_results} details the experimental setup and achieved results. Lastly, Section \ref{sec:conclusions} concludes the paper with the final remarks.

\section{Related Work} 
\label{sec:related}

The literature has presented several proposals that aim at treating or reducing the consequences of critical events in urban environments by using VANETs. Although many consider characteristics related to urban environments (e.g. traffic model, speed, and vehicle density), the impact of simultaneity (i.e. distributed over the same time span) and the mobility of critical events are not always taken into account during monitoring and information dissemination over the network.

The authors in~\cite{benkerdagh_cluster_2019} propose a new strategy for disseminating event alerts in VANETs based on optimization and clustering techniques. Instead of transmitting textual messages, the strategy uses codes to reduce the size of exchanged packets on the network. In addition, it ensures the stability of groups by using \emph{fitness} functions to select their members. However, the strategy is limited for considering only the occurrence of fixed events in the environment. In~\cite{siddiqua_icafe_2019}, a content-centric protocol called iCAFE was proposed, employing intelligent congestion avoidance and fast emergency services. It considers five types of packets for communication \textit{V2V} and \textit{V2I}, as well as a new congestion control mechanism. When a critical condition like a collision affects a vehicle, it sends emergency messages to the base stations present in the urban infrastructure to alert the nearest ambulance. Moreover, nearby vehicles are reported to leave the affected lane, so ambulances can reach the accident in time. However, \textit{iCAFE} does not apply cooperative monitoring, since only the affected vehicle transmits emergency messages.

In~\cite{Chou2017}, a system for the dissemination of emergency messages in vehicular networks is proposed, named \textit{Appropriate Vehicular Emergency Dissemination} (AVED). The objective is to transmit emergency messages of automobile accidents, where sensors embedded in the damaged vehicle send emergency messages to neighboring vehicles, which in turn pass them on to their own neighbors. AVED is based on the WAVE/DSR standard and maintains a periodically updated routing table, capable of determining the most reliable neighbor to forward the message over the network. However, the studies described up to now are limited to the occurrence of only fixed and non-simultaneous events. 

In~\cite{khakpour_using_2017}, the authors present an analysis of two clustering algorithms for tracking vehicles on transit roads. In the strategy applied by the algorithms, the groups are formed by vehicles that currently detect the target one and by those vehicles that have a high probability of detecting the target shortly thereafter. The group leader concentrates the collected data from all group members, analyzes in order to eliminate redundancies, and relays to the infrastructured network. Despite their effectiveness, both clustering algorithms centralize in the group leader the responsibility of deciding what to transmit to a control center. In~\cite{Derder2019}, a vehicle tracking protocol for VANETs was proposed. It consists of two phases executed asynchronously, i.e. \emph{discovery} (transmission of tracking request messages) and \emph{tracking} (detection and forwarding of packets to Road Side Units - RSUs). The authors define the concept of a virtual RSU, which is a vehicle responsible for relaying tracking request messages to the network. However, despite considering mobile events, the last two studies still do not take into account the occurrence of simultaneous ones. In a previous work~\cite{Andrade2020}, the MINUET system was proposed for a dynamic network monitoring and collaborative dissemination of critical events in urban VANETs. After detection, the system forms a network of grouped vehicles to monitor and disseminate the occurrence of simultaneous events, whether fixed or mobile. 

Table~\ref{tab:comparative} presents a comparative summary of the previously described works in terms of event and environment characteristics. From this comparison, we can see that only two works consider real traffic for proposal purposes, as in~\cite{Andrade2020} and \cite{Derder2019}. Also, the one in~\cite{Andrade2020} considers both fixed and mobile events, as well as their simultaneous occurrence. So, it is highlighted that most works do not take into account the complete set of characteristics, but the one described in~\cite{Andrade2020}. This way, the MINUET system has been chosen for the assessment proposed in this paper, since we aim to analyze the impacts of these characteristics in the monitoring and dissemination of information in urban VANET environments.

\begin{table}
\renewcommand{\arraystretch}{1.0}
	\centering
	\scriptsize
    \caption{Comparative summary of event and environmental characteristics.}
	\label{tab:comparative}
	\begin{tabular}{|c|| c | c | c | c | c |}
		\hline
		\centering
		\multirow{2}{*}{\textbf{
		RW 
		}} & \multicolumn{2}{c|}{\textbf{Event}} & \multicolumn{3}{c|}{\textbf{Environment}} \\
		\cline{2-6}
		& \textbf{Mobility} & \textbf{Simultaneity} & \textbf{Traffic model} & \textbf{Node speed} & \textbf{Density} \\
		\hline \hline
		\cite{benkerdagh_cluster_2019} & Fixed & No & Poisson traffic & variable &  variable \\
		\hline
		\cite{siddiqua_icafe_2019} & Fixed & No & Random Way-point & -- &  -- \\
		\hline
		\cite{Chou2017} & Fixed & No &  Freeway mobility &  variable &  variable \\
		\hline
	    \cite{khakpour_using_2017} & Mobile & No & Own traffic model & variable & variable \\
		\hline
		\cite{Derder2019} & Mobile & No & Manhattan traffic & constant & variable \\
		\hline
		\cite{Andrade2020} & Both & Yes & Luxembourg traffic & variable & variable \\
		\hline
	\end{tabular}
\end{table}

\section{The MINUET System}
\label{sec:minuet}

In this section, we describe the system's operation and present a model characterization for further assessment.

\subsection{System Description}

The MINUET\footnote{Available under request: \url{https://bitbucket.org/everaldoandrade/minuet}} (\textbf{M}onitor\textbf{IN}g and Dissemination of \textbf{U}rban \textbf{E}ven\textbf{T}s) \cite{Andrade2020} system aims to assist city authorities in the treatment of critical urban events through a dynamic monitoring and cooperative dissemination of video data streams, supported by the clustering of vehicles and an infrastructured network. MINUET adopts an adaptive and distributed coordination and control management, taking into account the temporal and spatial contexts of the vehicles in a \textit{crowdsensing} approach. It is based on two network services: 1) provides the event detection and maintenance of its monitoring, as well as video data dissemination; and 2) responsible for the creation and maintenance of a topology for passing on monitored information, established by using vehicle clustering techniques. MINUET operates on an urban environment where vehicles communicate with each other (V2V) and with the infrastructure (V2I) through Base Stations (BSs). In such environments, critical events may occur randomly both in time and space and they can change its position over time. This situation imposes additional challenges to the monitoring of events in dynamic networks (VANETs), since we also need to cope with the mobility of the event.

In this context, the urban environment model is composed of a set $V$ of $n$ vehicles (nodes) denoted by $\{v_{1}, v_{2},...,v_{n}\}$ and a set $E$ of $m$ events denoted by $\{ev_1, ev_2,...,ev_m\}$, where any vehicle $v_i \in V$ can detect a event $ev_j \in E$ if $ev_j$ is within the range of $v_i$. Every event $ev_j$ has a lifetime that begins at the moment the event occurs $t_{ev_j, 0}$ until it vanishes or is handled by an acknowledged authority $t_{ev_j, f}$. The event can also move over time, arising at position $s_{ev_j, 0}$ and vanishing at position $s_{ev_j, f}$. Therefore, the tuple $L_{ev_j} = \langle s_{ev_j}, t_{ev_j} \rangle$ denotes that the event $ev_j$ is at position $s_{ev_j}$ at time $t_{ev_j}$, where $s_{ev_j, 0} \leq s_{ev_j} \leq s_{ev_j, f}$ and $t_{ev_j, 0} \leq t_{ev_j} \leq t_{ev_j, f}$. If $ev_j$ is fixed, then $s_{ev_j}$ is constant over time. Once detected, the event must be monitored with the support of a cluster of vehicles. So, the set of monitoring vehicles that detect an event $ev_j$ is defined as $C^k(L_{ev_j}) = \{\forall v_d \ \ | \ \ v_d \ \ \text{monitors} \ \ ev_j \ \ \text{in} \ \ L_{ev_j} \ \ \text{and} \ \ v_d \in V\}$, since each cluster of vehicles depends on the space-time characteristics of the event. Then, for a specific event $ev_j$, the urban model can be seen as several monitoring clusters formed over space and time, denoted by $\{ C^k(L_{ev_j}) \ \ | \ \ k = 1, 2, ..., K\}$.



\begin{figure}[b]
\centering
\includegraphics[draft=false,width=1.0\columnwidth]{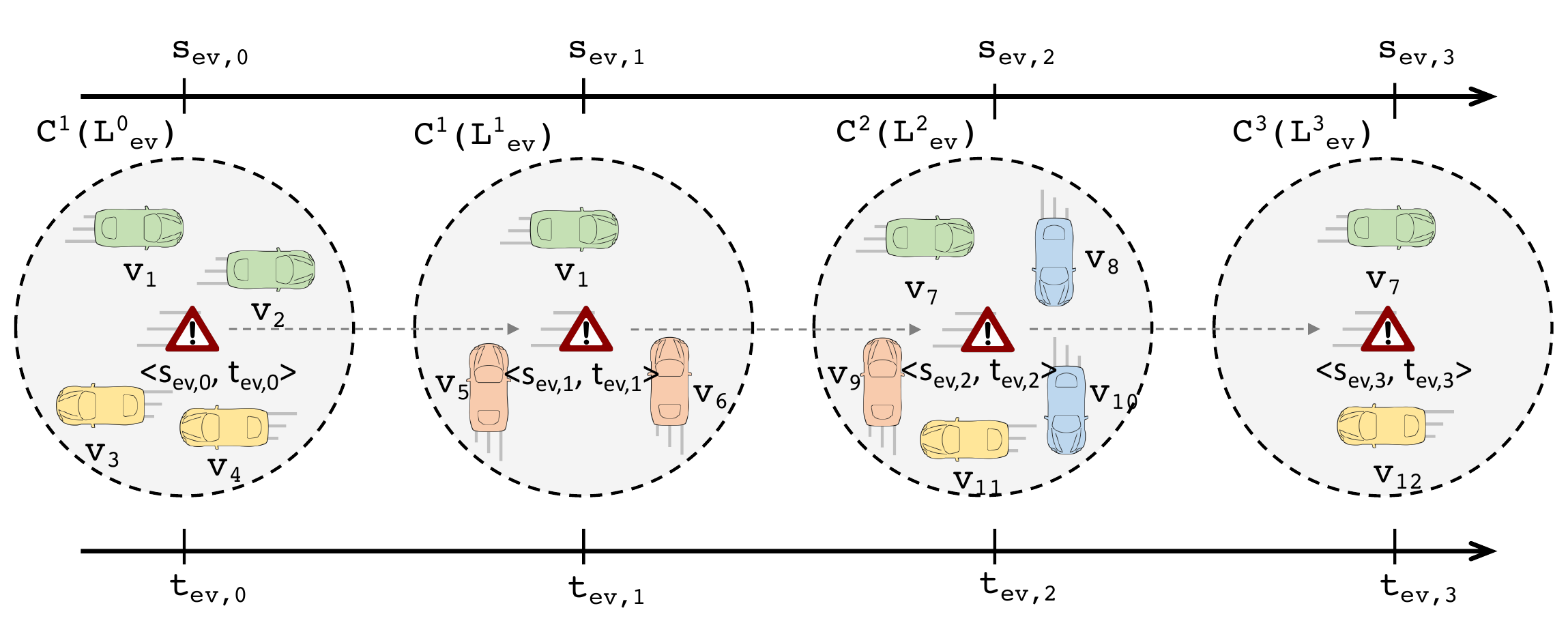}
\caption{Monitoring urban environment for a mobile event.}
\label{fig:mobile}
\end{figure}

Fig. \ref{fig:mobile} illustrates the urban environment where MINUET operates. For the sake of simplification, it shows an unique event $ev$ appearing in position $s_{ev, 0}$ at time $t_{ev, 0}$ and moving to position $s_{ev, 3}$ at time $t_{ev, 3}$, when it ceases ($t_{ev, 3} = t_{ev, f}$). MINUET supports the formation or maintenance of clusters to monitor $ev$ at each $L_{ev}$. As the event moves, clusters are formed to monitor it over a specific time interval. 
In the figure, there are three clusters monitoring the event throughout its duration (from $t_{ev, 0}$ to $t_{ev, 3}$). Cluster $C^1(\cdot)$ monitors $ev$ from $L_{ev}^0 = \langle s_{ev, 0}, t_{ev, 0} \rangle$ to $L_{ev}^1 = \langle s_{ev, 1}, t_{ev, 1} \rangle$, while $C^2(L_{ev}^2)$ and $C^3(L_{ev}^3)$ monitors $ev$ at $L_{ev}^2 = \langle s_{ev, 2}, t_{ev, 2} \rangle$ and $L_{ev}^3 = \langle s_{ev, 3}, t_{ev, 3} \rangle$, respectively. Since the event moves in the urban environment comprising of diverse traffic lane topologies, each cluster can have vehicles with different orientations. It is worth noting that a vehicle can be a member of the same cluster or of distinct clusters at different times. For instance, vehicle $v_1$ is member of the first cluster at two different moments ($L_{ev}^0$ and $L_{ev}^1$), while $v_7$ is member of second and third clusters at $L_{ev}^2$ and $L_{ev}^3$, respectively.

\begin{figure}
\centering
\includegraphics[draft=false,width=1.0\columnwidth]{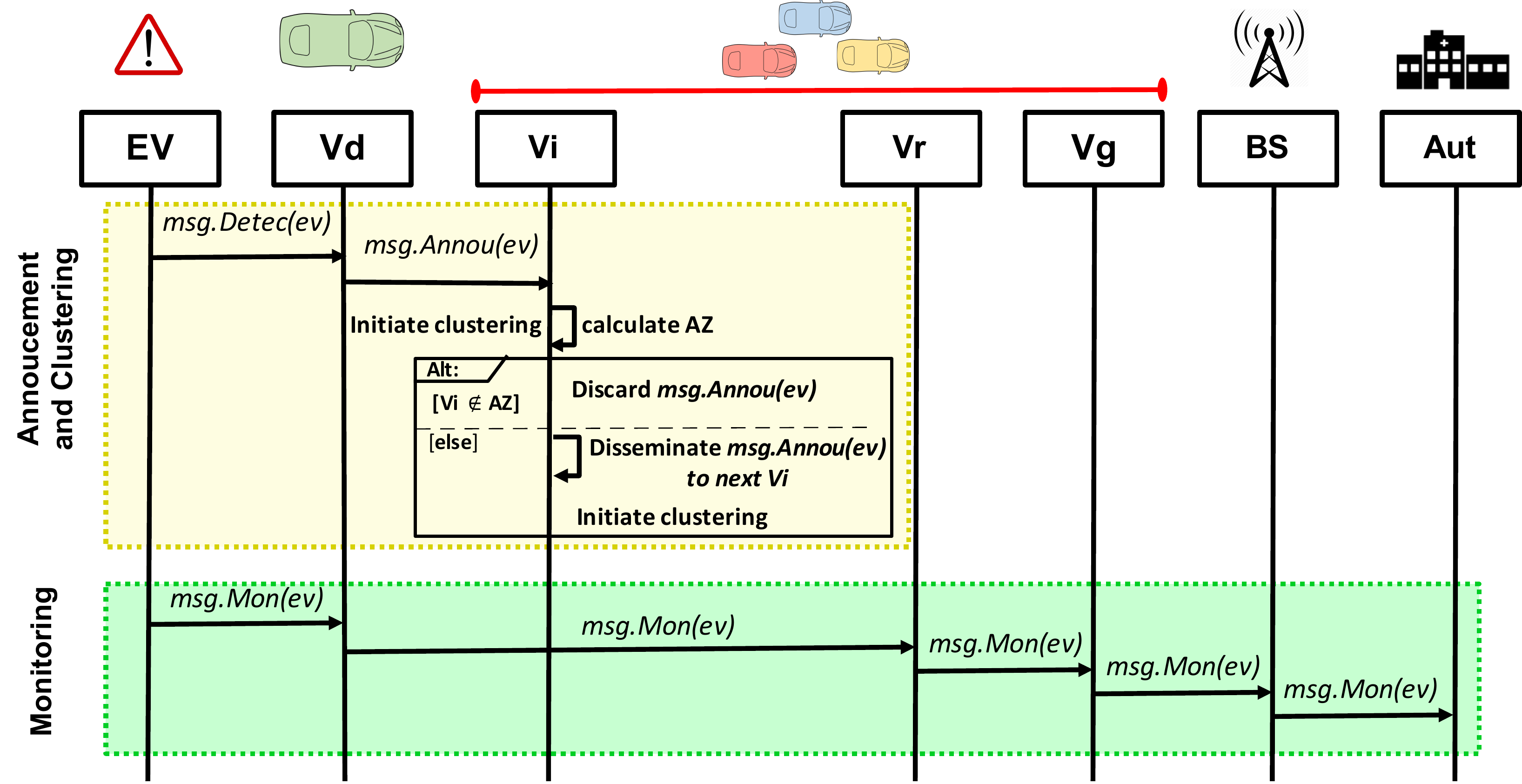}

\caption{The exchanged messages in the MINUET system.}
\label{fig:fluxo}
\end{figure}

Fig.~\ref{fig:fluxo} depicts a diagram representing the interactions among MINUET participants. We denote $v_d$ as a detecting vehicle, $v_i$ as any other intermediary vehicle, $v_r$ as a relay vehicle, $v_g$ as a gateway, and $BS$ as a base station. The interaction occurs by the exchanging of messages during the \textbf{announcement}, \textbf{clustering}, and \textbf{monitoring} phases. When detecting event $ev$ via message \textit{msg.Detec(ev)}, $v_d$ sends announcement messages \textit{msg.Annou(ev)} to its one-hop neighbours. In line to avoid transmitting a message indefinitely, MINUET employs an \textit{announcement zone} $AZ$ that is limited by the maximum dissemination time $t_{max}$, advertised in the announcement message. Upon receiving \textit{msg.Annou(ev)}, $v_i$ verifies whether $t_{max}$ has been exceeded. If so, $v_i$ is not in the $AZ$ and discards \textit{msg.Annou(ev)}. Otherwise, the announcement message is disseminated to the vehicles one-hop away from $v_i$, repeating the $AZ$ calculation process. It continues until \textit{msg.Annou(ev)} reaches vehicles that are not in the $AZ$ coverage. 

Inside the $AZ$, $v_d$ and $v_i$ start the clustering process in the search of collaboration at monitoring and disseminating $ev$ information. Each participating member defines its role within the group, which can be that of a \textit{monitor} ($v_d$), \textit{relay} ($v_r$), or \textit{gateway} ($v_g$). While detecting $ev$, $v_d$ remains keeping track of the event and sending monitoring messages \textit{msg.Mon(ev)}, which are disseminated through $v_r$ until reach $v_g$. This last gateway vehicle is responsible for delivering the monitoring messages to the base station $BS$ and, consequently, to the corresponding authority represented by $Aut$. We presume that $Aut$ will take the appropriate measures in the treatment of event $ev$. Depending on the network topology in a given moment, a single vehicle can play more than one role in the system.

\begin{figure}[h]
\centering
\includegraphics[draft=false,width=0.9\columnwidth]{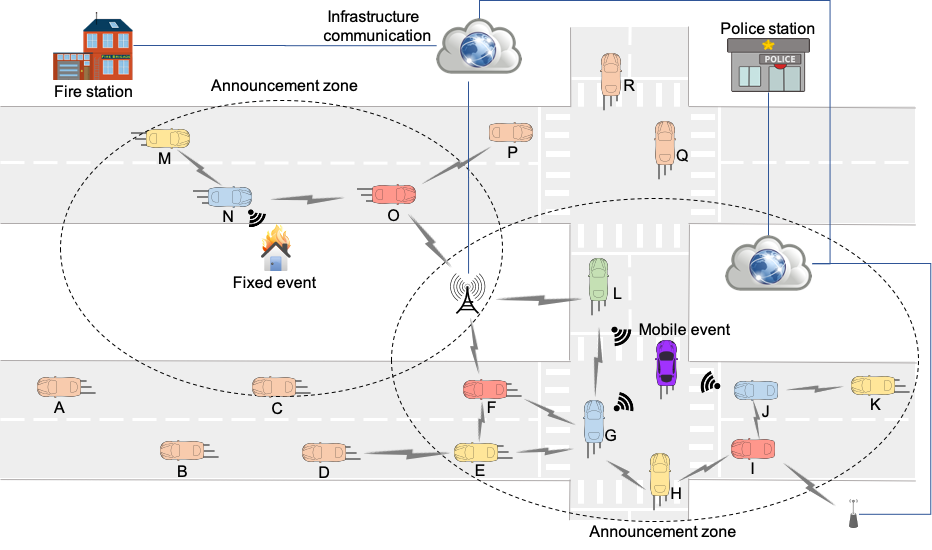}
\caption{A MINUET operation example.}
\label{fig:exemplo}
\end{figure}

Fig.~\ref{fig:exemplo} exemplifies the MINUET operations in a urban environment where a set of collaborative vehicles detect and transmit visual information of critical events to the base stations in a given time period. This scenario comprises, particularly, 19 vehicles represented by letters ($A$, $B$, $C$, ..., $R$), a fixed event (fire) and a mobile event (car on the run). In a given moment, vehicles $G$, $J$ and $L$ detect the mobile event while vehicle $N$ detects the fixed event, thus playing the role of monitoring vehicles. They collect information about the contexts of the events and disseminate this information to their neighbours via announcement messages. Based on $t_{max}$, vehicles verify if they belong to the $AZ$ and then participate in the clustering. Formed groups are comprised of member nodes and a leader, which is defined according to the clustering algorithm. In this example, when $D$ and $P$ receive \textit{msg.Annou(ev)}, they certify not belonging to the $AZ$ and discard participation in the clustering. As long as the monitoring vehicles are capable of observing the events, they continue announcing to the other vehicles and new groups are collaboratively formed.

The same moment the monitoring vehicles detect the events they initiate a continuous monitoring process, in which multimedia data is disseminated in the vehicular network. Collaboratively, members of the group pass the data on to their neighbours, thus becoming relay vehicles ($E$, $H$, $K$ and $M$). If a vehicle is in the reach of a base station, it will be able to deliver the monitoring data, thus playing the role of gateways ($F$, $I$, $L$ and $O$). In our particular case, $L$ not only monitors the mobile event but also delivers the information to a close base station. Finally, the corresponding authorities receive the data concerning the events they are in charge of. The authorities then recognize and take the appropriate measures to address the events.

\subsection{Evaluation System Model}

In order to measure the efficiency in disseminating videos under different critical events' contexts, we model some metrics to quantify important characteristics of the environment under the operation of the MINUET system. We first consider the set of all vehicles in the system at a given time $t$ by $N(t)$. Moreover, $n_{v_d}(t)$ represents the system's capacity of detecting an event, denoted by the amount of vehicles that detect the event in the instant of time $t$, from all the ones in the $AZ$:

\begin{align}\label{eq:num_detecting_vehicles}
n_{v_d}(t) = \sum_{v_i \in N(t)} d_i(t); \text{\space with\space} 
d_i(t)=  \begin{cases}
    1, & \text{if $v_i$ detects the event at $t$} \\
    0, & \text{otherwise.}
            \end{cases}
\end{align}

We also assume that $n_{v_c}(t)$ represents the number of collaborating vehicles inside the $AZ$, i.e. vehicles either monitoring or disseminating information about $EV$ at time $t$. It shows the collaboration degree in MINUET and is given by:

\begin{align}\label{eq:num_collaborating_vehicles}
n_{v_c}(t) = \sum_{v_i \in N(t)} c_i(t); \text{\space with\space}
c_i(t)=  \begin{cases}
    1, & \text{if $v_i$ cooperates at time $t$} \\
    0, & \text{otherwise.}
            \end{cases}
\end{align}

To highlight the overhead imposed by MINUET, we define the two types of messages employed in the system as: clustering packets ($CP$) and monitoring packets ($MP$). Then, the communication \emph{overhead} is assumed to be the amount of \textit{generated} packets from these 2 types of above-mentioned messages, given at the moment of time $t$ as: $CP_g(t)$, for clustering messages; and $MP_g(t)$, for monitoring messages. Also of particular interest, the number of monitoring packets \textit{received} by all BSs at time $t$ is given by $MP_{r}(t)$. These last three metrics are expressed respectively by:

\begin{equation}\label{eq:num_msgs}
\def\arraystretch{1.3}
\def\arraycolsep{10pt}
\begin{array}{@{}ll@{}}
CP_{g}(t) = \displaystyle\sum_{v_k \in N(t)} g_k^c(t); \text{\space(a)} & MP_{g}(t) = \displaystyle\sum_{v_k \in N(t)} g_k^m(t); \text{\space(b)}  \\
& \\
MP_{r}(t) = \displaystyle\sum_{j \in BS} r_j^m(t); \text{\space(c)} &  \\
\end{array}
\end{equation}

\noindent
where $g_k^{(\cdot)}(t)$ is the number of each type of generated packets at time $t$, given that $v_k$ is a detecting or cooperating vehicle; and $r_j^m(t)$ is the number of monitoring packets received by BS $j$ at time $t$. We highlight that Eq.~\ref{eq:num_msgs}(a) and \ref{eq:num_msgs}(b) accounts for the amount of generated packets by the MINUET system, which is considered as the communication overhead imposed by clustering and monitoring phases. Besides, Eq.~\ref{eq:num_msgs}(b) and \ref{eq:num_msgs}(c) represent respectively the system's capacity to monitor and deliver information to the corresponding authorities.

As Eq.~\ref{eq:num_msgs}(c) accounts for all monitoring packets delivered to the base stations at a given time $t$, we assume that a single packet can be received in more than one BS. So, it is possible the replication of packets in different BSs. Let us represent this by assume that, at a given time $t$, the number of received packets in a BS $j$ is the sum of packets received only (\textit{singly}) by that BS and the redundant packets also received by the same BS, namely: $r_j^m(t) = r_{j,sin}^m(t) + r_{j,red}^m(t)$.

The level of redundancy in the delivery of monitoring packets provided by the system is also a desired measure, i.e. the number of repeated packets delivered at the base stations. Let us assume that in a time interval $\Delta t$, the indication of redundant packets in the system is the ratio of duplicated packets by the total number of received ones, denoted as $\mathcal{R}(\Delta t)$. Likewise, to express the amount of non-redundant packets delivered to the BSs during a specific time interval we define $\mathcal{S}(\Delta t)$. Both metrics are defined as:

\begin{equation}\label{eq:redundancy}
\def\arraystretch{1.3}
\def\arraycolsep{4pt}
\begin{array}{@{}ll@{}}
\mathcal{R}(\Delta t) = \dfrac{\displaystyle\int_{\Delta t}\sum_{j \in BS} r_{j,red}^m(t)dt}{\displaystyle\int_{\Delta t} \sum_{j \in BS} r_j^m(t)dt}; \text{\space(a)} & \mathcal{S}(\Delta t) = \dfrac{\displaystyle\int_{\Delta t}\sum_{j \in BS} r_{j,sin}^m(t)dt}{\displaystyle\int_{\Delta t} \sum_{j \in BS} r_j^m(t)dt}. \text{\space(b)}  \\
\end{array}
\end{equation}


Another important metric is the average delay of deliverd monitoring packets. Let us identify $P_r(\Delta t)$ as the set of packets received by all BSs in a $\Delta t$ time interval; $t_{p,g}$ and $t_{p,r}$ as the times a packet $p$ is generated and received, respectively. The average delay is denoted by $\mathcal{D}_{avg}(\Delta t)$ and expressed by:

\begin{align}\label{eq:avg_delay}
\mathcal{D}_{avg}(\Delta t) = \frac{\displaystyle\sum_{p \in P_r(\Delta t)} (t_{p,r} - t_{p,g})}{\displaystyle\int_{\Delta t} \sum_{j \in BS} r_j^m(t)dt}.
\end{align}

In order to evaluate the overhead imposed by any applied clustering technique, we define $\mathcal{C}(\Delta t)$ as the ratio of clustering messages in comparison to all 
generated messages in the system during $\Delta t$ time interval, given by:

\begin{equation}\label{eq:ratio_clustering_msgs}
\mathcal{C}(\Delta t) = \dfrac{\displaystyle\int_{\Delta t}\displaystyle\sum_{v_k \in N(t)} g_k^c(t)dt}{\displaystyle\int_{\Delta t}\displaystyle\sum_{v_k \in N(t)} \bigg(g_k^a(t) + g_k^c(t) + g_k^m(t)\bigg)dt}, 
\end{equation}

\noindent
where, again, $g_k^{(\cdot)}(t)$ is the corresponding number of generated packets if $v_k$ is sending clustering or monitoring messages at time $t$. 

Considering that any vehicle transmitting clustering packets belongs to at least one group, we define the ratio of clustering vehicles in a given time interval as $\mathcal{G}(\Delta t)$, based on the relation between the number of those transmitting clustering packets and the total number of vehicles in the system:

\begin{align}\label{eq:ratio_clustering_vehicles}
    \mathcal{G}(\Delta t) = \displaystyle\dfrac{ \displaystyle\int_{\Delta t}\sum_{v_k \in N(t)} I\Big(g_k^c(t)\Big)dt }{ \displaystyle\int_{\Delta t}\sum_{x = 1}^{|N(t)|}x dt }; \text{with\space} 
I\Big(g_k^c(t)\Big)=  \begin{cases}
    1, & \text{if $g_k^c(t) > 0$} \\
    0, & \text{otherwise.}
            \end{cases}
\end{align}

Finally, the number of \textit{formed groups} by the corresponding clustering technique is expressed as $\mathcal{F}(\Delta t)$, given by:

\begin{align}\label{eq:formed_groups}
\mathcal{F}(\Delta t) = \displaystyle\int_{\Delta t}\sum_{v_i \in N(t)} \mathcal{U}\Big(G_{ID}(v_i) \Big) dt,
\end{align}

\noindent
where $G_{ID}(\cdot)$ gives the group ID of a particular vehicle; and $\mathcal{U}(G_{ID})$ is a \textit{unique function} which results $1$ when is the first time $G_{ID}$ appears in the system and $0$ otherwise. Table \ref{tab:metricas_aval} summarizes the derived notation. 

\begin{table}[h]
	\renewcommand{\arraystretch}{1.4}
	\centering
\footnotesize
	\caption{Notation summary applied by equations.}
	\label{tab:metricas_aval}
	\begin{tabular}{| m{9em} || m{25em} |}
		\hline
	{\bf Equation} & {\bf Description} \\
	\hline \hline
	 $n_{v_{d}}(t)$ \ \ (Eq. 1) & The total number of detecting vehicles at time $t$. \\  
		\hline
		$n_{v_{c}}(t)$ \ \ (Eq. 2) & The total number of collaborating vehicles at time $t$. \\
        \hline
		$CP_g(t)$ \ \ (Eq. 3(a)) & The total number of clustering packets generated at $t$. \\
		\hline
		$MP_g(t)$ \ \ (Eq. 3(b)) & The total number of monitoring packets generated at $t$. \\
		\hline
		$MP_r(t)$ \ \ (Eq. 3(c)) & The total number of monitoring packets received at $t$. \\
		\hline
        $\mathcal{R}(\Delta t)$ \ \ (Eq. 4(a)) & The ratio of redundant monitoring packets. \\
        \hline
		$\mathcal{S}(\Delta t)$ \ \ (Eq. 4(b)) & The ratio of non-redundant monitoring packets. \\
        \hline
		$\mathcal{D}_{avg}(\Delta t)$ \ \ (Eq. 5) & The average delay in the system. \\
		\hline        
    	$\mathcal{C}(\Delta t)$ \ \ (Eq. 6) & The ratio of clustering packets. \\
        \hline
    	$\mathcal{G}(\Delta t)$ \ \ (Eq. 7) & The ratio of clustering vehicles. \\
        \hline
		$\mathcal{F}(\Delta t)$ \ \ (Eq. 8) & The number of formed groups by the clustering technique. \\
		\hline
	\end{tabular}
\end{table}

\begin{figure*}[!htp]
\centering
\includegraphics[draft=false,scale=0.46]{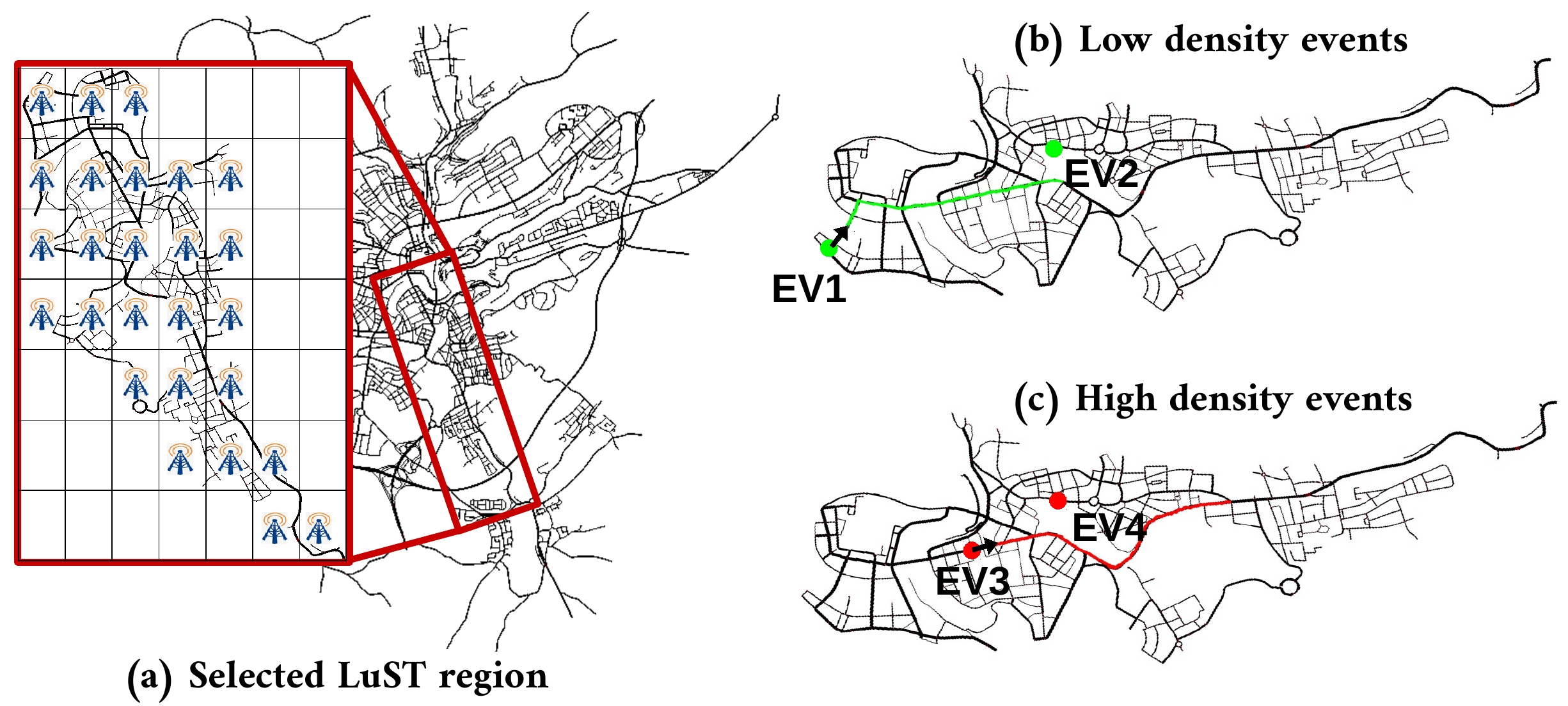}
\caption{LuST urban traffic environment and distribution of events.
}
\label{fig:events_distribution}
\end{figure*}

\section{Experimental Setup and Results} \label{sec:setup_results}

This section describes the evaluation scenarios and obtained results in terms of the proposed metrics. The assessment is carried out through simulations, where realistic models of urban environments are considered.

\subsection{Implementation and Experimentation Environment}

We implemented MINUET in C++ programming language and the simulation was carried out using NS-3 version 3.28, along with SUMO. We used IEEE 812.11p with 5.9GHz frequency band, 10MHz bandwidth and bitrate of 6Mbps in the simulation.
Moreover, we assume two vehicle clustering algorithms: \textit{DCA}~\cite{Tal2016} and \textit{PCTT}~\cite{khakpour_using_2017}. 
The evaluation environment consists of a region of the LuST project (\textit{Luxembourg SUMO Traffic})\footnote{Version 2.0 available at:~\url{https://github.com/lcodeca/LuSTScenario}}, 
which uses 
the road network of the city of Luxembourg, with real urban traffic patterns over a period of 24 hours. Vehicular density varies over time, as with realistic 
scenarios. 
The positions of the fixed and mobile events and the routes of the mobile events were defined beforehand.
Vehicles at most 10 meters away from the events can detect and monitor them. 
For our simulations, we used an Intel Core i7 machine with 16GB RAM and 
Ubuntu OS 18.04.4 LTS.


Fig.~\ref{fig:events_distribution}(a) represents the urban model of approximately $6.5km^2$ applied in our simulations. The area was divided into quadratic sub-areas with base stations uniformly arranged in order to cover the entire urban infrastructure. The assessment takes into account the occurrence of multiple events of different characteristics, distributed in time and physical space. Four events lasting five minutes each were distributed over time and space in the scenario, representing different traffic contexts. Fig.~\ref{fig:events_distribution}(b) and Fig.~\ref{fig:events_distribution}(c) depict the distribution of the two types of events, the mobile ones (\textit{EV1} and \textit{EV3}) and the fixed ones (\textit{EV2} and \textit{EV4}); under low-density (6:28-6:33am) and high-density (8:20-8:25am) time scenarios. The mobile events move through the roads they arise, with the arrows' direction indicating the pathway. Table~\ref{tab:setup_parameters} presents the setup parameters used in the simulations.


\begin{table}[ht]
	\centering
	\footnotesize
	\caption{Setup parameters applied at the simulations.}
	\label{tab:setup_parameters}
	\vspace{-2mm}
	\begin{tabular}{| m{10em} || C{4em} | C{4em} | C{4em} | C{4em} |}
		\hline
		\centering
		\textbf{PARAMETER} & \textbf{EV1} & \textbf{EV2} & \textbf{EV3} & \textbf{EV4} \\ 
		\hline \hline
		Simulation time ($\Delta t$) & \multicolumn{4}{c|}{6 min} \\
		\hline
		Simulated area & \multicolumn{4}{c|}{$\approx$ $6.5Km^2$} \\
		\hline
		Event duration & \multicolumn{4}{c|}{5 min} \\
		\hline
		Fixed event & No & Yes & No & Yes \\
		\hline
		Position & -- & (7682, 5878.20) & -- & (7682, 5878.20) \\
		\hline
		Time intervals & \multicolumn{2}{c|}{[6:28am, 6:33am]} & \multicolumn{2}{c|}{[8:20pm, 8:25pm]} \\
		\hline
		Transmission range & \multicolumn{4}{c|}{$\approx$200m} \\
		\hline
		Transport Protocol & \multicolumn{4}{c|}{UDP} \\
		\hline
		PHY/MAC & \multicolumn{4}{c|}{IEEE 802.11p} \\
		\hline
		No. of base stations & \multicolumn{4}{c|}{26} \\
		\hline
		Propagation model & \multicolumn{4}{c|}{YANS} \\
		\hline
	\end{tabular}
\end{table}

\subsection{Analysis of Performance Results}

\begin{figure}[ht]
\centering
\subfloat{\includegraphics[draft=false,width=0.32\columnwidth]{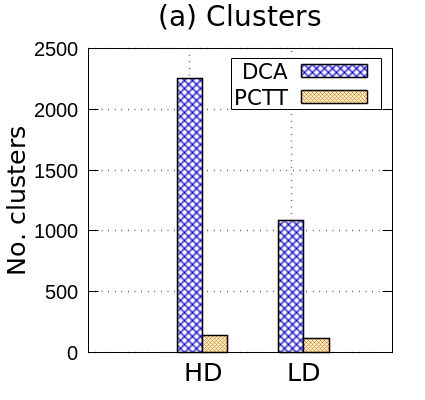}}
\subfloat{\includegraphics[draft=false,width=0.32\columnwidth]{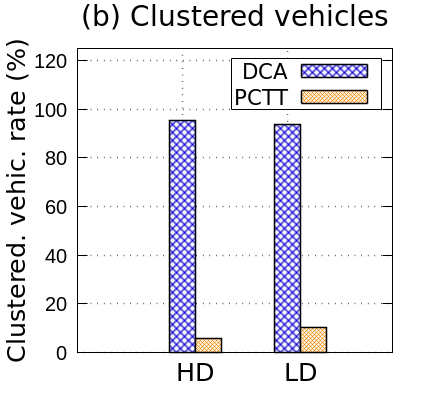}}
\subfloat{\includegraphics[draft=false,width=0.32\columnwidth]{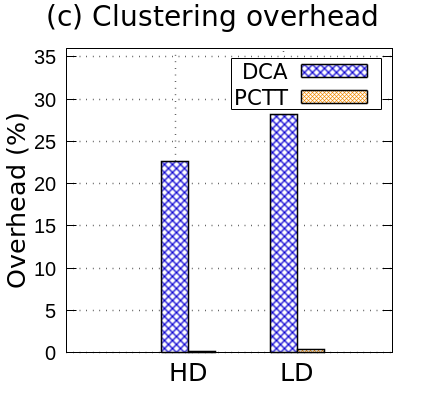}}
\caption{Clustering algorithms data (Eq. \ref{eq:formed_groups}, \ref{eq:ratio_clustering_vehicles} and \ref{eq:ratio_clustering_msgs}).}
\vspace{-0.3cm}
\label{fig:nfg_tva_oa}
\end{figure}

Firstly, we aim to show how the two clustering algorithms applied in our evaluation behave under the operations of the MINUET system. The general results shown in Fig.~\ref{fig:nfg_tva_oa} helps the reader to understand how the characteristics of the DCA and PCTT impact on further results. Fig.~\ref{fig:nfg_tva_oa}(a) shows the number of formed groups (Eq. \ref{eq:formed_groups}) by each algorithm, and is evident that DCA forms a much higher number of groups than PCTT. This happens because DCA presents a proactive grouping strategy, i.e. multiple groups are formed during monitoring and dissemination. Likewise, Fig.~\ref{fig:nfg_tva_oa}(b) shows that the formed groups are also much larger with DCA, since the percentage of grouped vehicles is higher (derived from Eq. \ref{eq:ratio_clustering_vehicles}) than with PCTT. Lastly, the overhead of clustering packets (Eq. \ref{eq:ratio_clustering_msgs}) can be seen in Fig.~\ref{fig:nfg_tva_oa}(c) and, as expected, the values are lower for the PCTT. In summary, regardless of which density scenario, DCA forms a higher number of and larger groups due to its inherent behavior, but it incurs more overhead because of that.


\begin{figure}[!htb]
\centering
\subfloat{\includegraphics[draft=false,width=0.4\columnwidth]{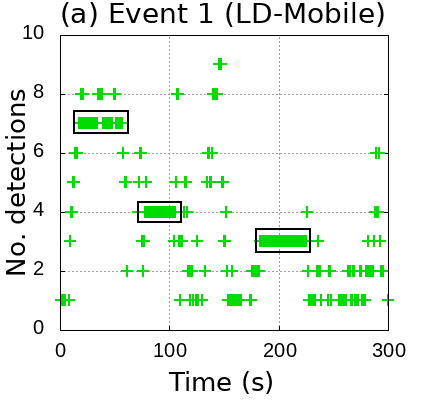}}
\subfloat{\includegraphics[draft=false,width=0.4\columnwidth]{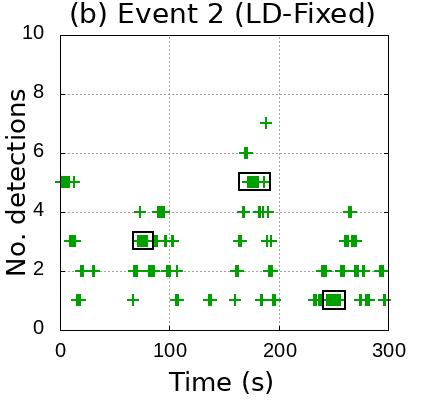}} \\
\subfloat{\includegraphics[draft=false,width=0.4\columnwidth]{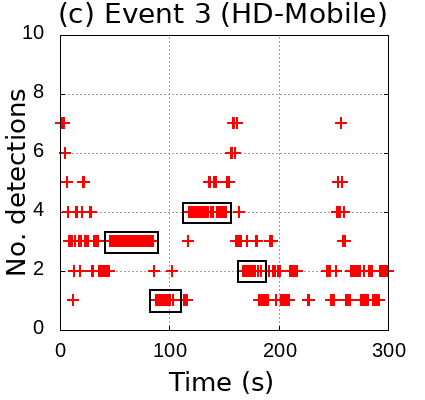}}
\subfloat{\includegraphics[draft=false,width=0.4\columnwidth]{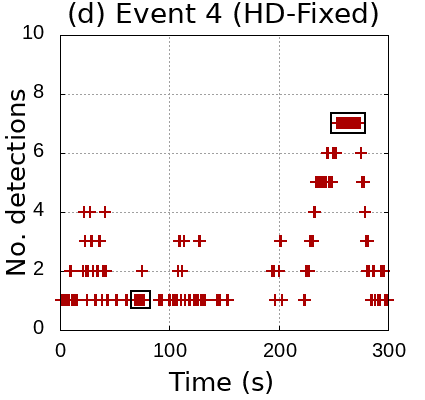}}
\caption{Number of vehicles detecting both mobile and fixed events, and at low and high density scenarios (Eq. \ref{eq:num_detecting_vehicles}).}
\vspace{-0.6cm}
\label{fig:nvd}
\end{figure}

The graphs in Fig.~\ref{fig:nvd} show the system's ability to detect and, consequently, monitor events over time (Eq. \ref{eq:num_detecting_vehicles}) in different traffic contexts. Mobile and fixed events at the high-density (HD) scenarios (Fig.~\ref{fig:nvd}(c) and Fig.~\ref{fig:nvd}(d)) presented more continuous monitoring in relation to low-density (LD) ones (Fig.~\ref{fig:nvd}(a) and Fig.\ref{fig:nvd}(b)). This is expected due to the fact that high-density environments have a larger number of vehicles closer to the event and for a longer period of time; especially for mobile events that move in the same direction of the traffic (likewise our experiments). Still, in the low-density scenarios, a high number of detections in shorter intervals of time is perceived. Although all events were monitored, Fig.~\ref{fig:nvd}(a) and Fig.~\ref{fig:nvd}(c) show that mobile events are perceived longer, implying a monitoring period of approximately 87\% and 88\% of the total elapsed time at the low and high density scenarios, respectively; against 51.5\% and 56\% of fixed events (\textit{EV2} and \textit{EV4}).
For the mobile events, there are periods where the events are continuously detected, indicated by the rectangles on the graphs. In such cases, vehicles holding the same orientation as the mobile event perform most of the detections at different times. Thus, the vehicles have higher probability to deliver the event data since they
they monitor the event for longer periods. This is not the case for fixed events where vehicles lost track of the events usually after a short window of time. These results show that monitoring approaches that take into account the mobile event orientation favor the detection and dissemination.

\begin{figure}[h]
\centering
\subfloat{\includegraphics[draft=false,width=0.4\columnwidth]{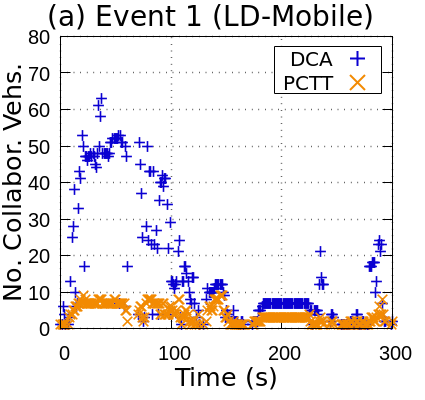}}
\subfloat{\includegraphics[draft=false,width=0.4\columnwidth]{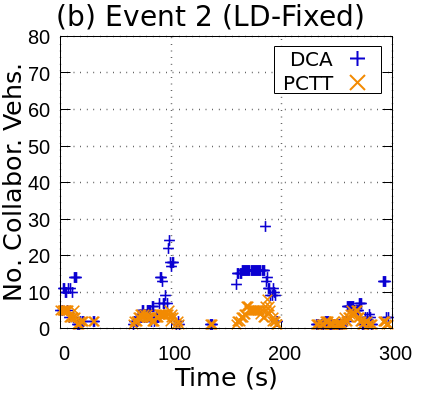}} \\
\subfloat{\includegraphics[draft=false,width=0.4\columnwidth]{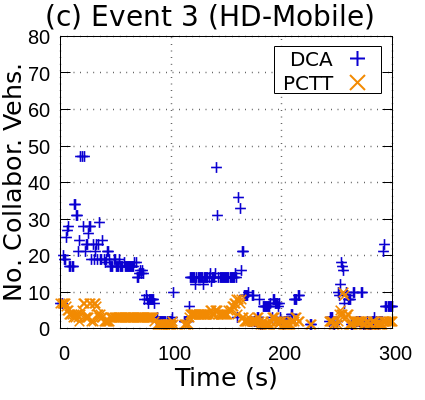}}
\subfloat{\includegraphics[draft=false,width=0.4\columnwidth]{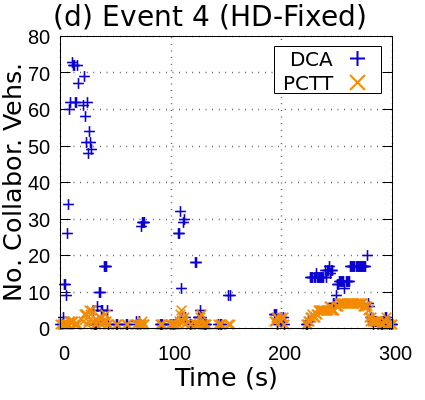}}
\caption{Number of collaborative vehicles disseminating mobile and fixed events at low and high density scenarios (Eq. \ref{eq:num_collaborating_vehicles}).}
\label{fig:nvc}
\end{figure}

The cooperative capacity of vehicles during the monitoring and dissemination of mobile and fixed events is confirmed by the number of collaborating vehicles per time in Fig.~\ref{fig:nvc} (Eq. \ref{eq:num_collaborating_vehicles}). The system presented a \textit{crowdsensing} collaboration using both DCA and PCTT, for all density scenarios. For mobile events (\textit{EV1} and \textit{EV3}), there is a higher number of vehicles collaborating over time, either monitoring or disseminating. 
This is also explained by the fact that vehicles holding the same orientation as the mobile event keep collaborating for longer periods of time.
The figure also indicates that a high density does not always imply higher cooperation in urban environments.
This is highlighted in the first 100 seconds of Fig.~\ref{fig:nvc}(a) and Fig.~\ref{fig:nvc}(c), in which \textit{EV1} showed a higher number of cooperating vehicles than \textit{EV3}. Further, a higher collaboration is achieved when using the DCA in relation to PCTT, mainly when monitoring mobile events. This behaviour is due to the higher number (Fig.~\ref{fig:nfg_tva_oa}(a)) and larger (Fig.~\ref{fig:nfg_tva_oa}(b)) groups formed by DCA and its proactive strategy.


\begin{figure}[!htb]
\centering
\subfloat{\includegraphics[draft=false,width=0.8\linewidth]{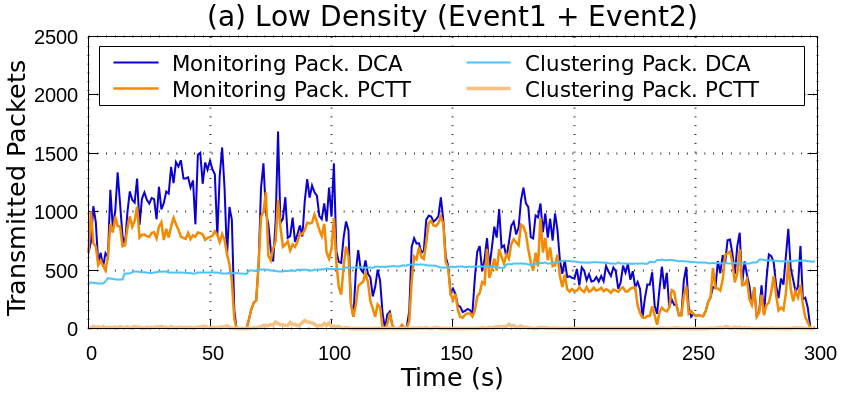}}\\
\subfloat{\includegraphics[draft=false,width=0.8\linewidth]{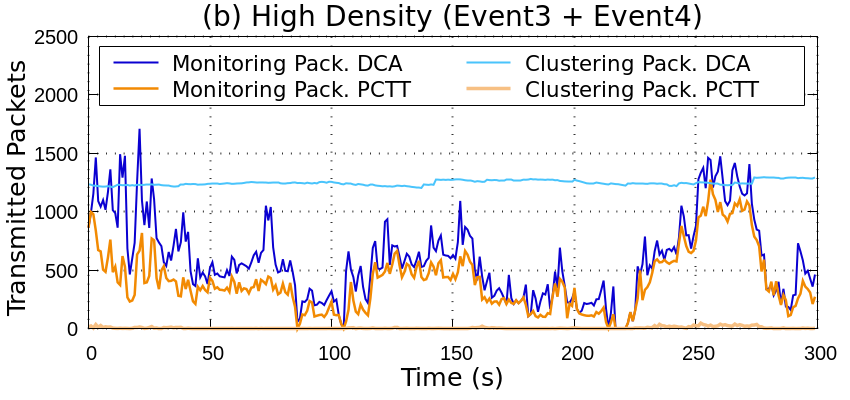}}
\caption{The communication overhead by generated monitoring and clustering packets (Eq. \ref{eq:num_msgs}(a) and \ref{eq:num_msgs}(b)).}
\label{fig:opr}
\end{figure}

The communication \textit{overhead} is evidenced in Fig.~\ref{fig:opr} by the number of generated packets during the occurrence of events, namely the monitoring and clustering packets incurred in the network (Eq. \ref{eq:num_msgs}(a), \ref{eq:num_msgs}(b) and \ref{eq:num_msgs}(c)). Even considering that events occur simultaneously and at different traffic contexts, the system presented a constant flow of clustering packets and a continuous flow of monitoring packets most of the time. According to Fig.~\ref{fig:opr}(a) and Fig.~\ref{fig:opr}(b), the amount of monitoring packets transmitted is slightly larger when using DCA, for both density scenarios. Also, the number of clustering packets transmitted by PCTT is lower compared to DCA, as confirmed by the total number of formed groups in Fig.~\ref{fig:nfg_tva_oa}(c). This happens because the PCTT considers only vehicles that have detected the event at least once as able to transmit clustering packets. Thus, under the operations of the MINUET, we can recognize the management of dynamic monitoring and distributed dissemination of information, taking into account different traffic contexts and event characteristics.


\begin{figure}
\centering
\subfloat{\includegraphics[draft=false,width=0.76\linewidth]{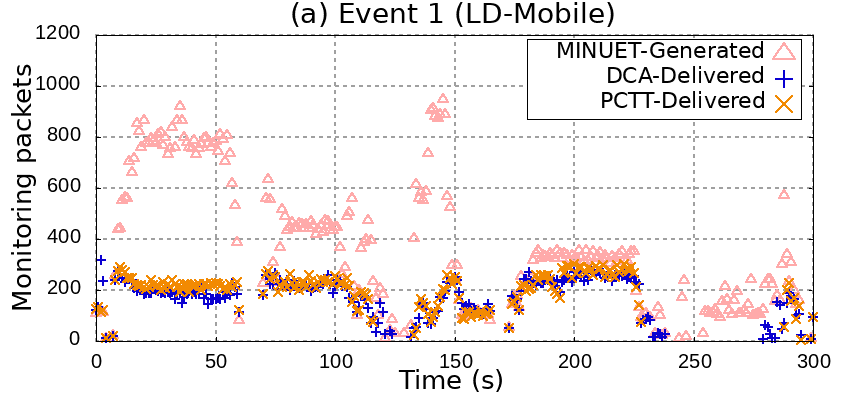}}\\
\subfloat{\includegraphics[draft=false,width=0.76\linewidth]{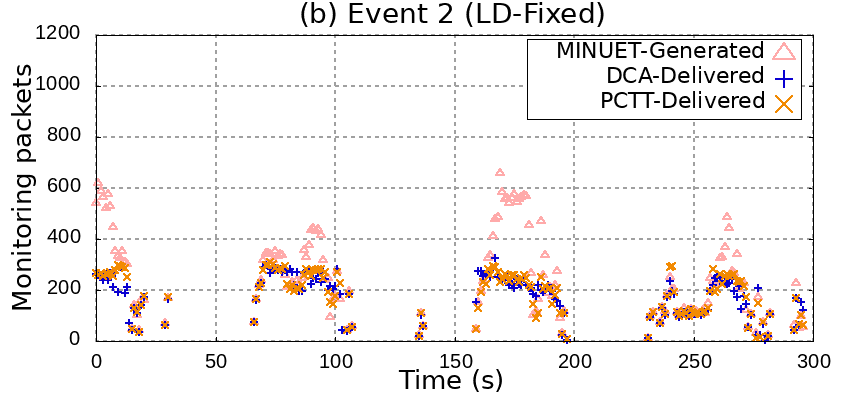}}\\
\subfloat{\includegraphics[draft=false,width=0.76\linewidth]{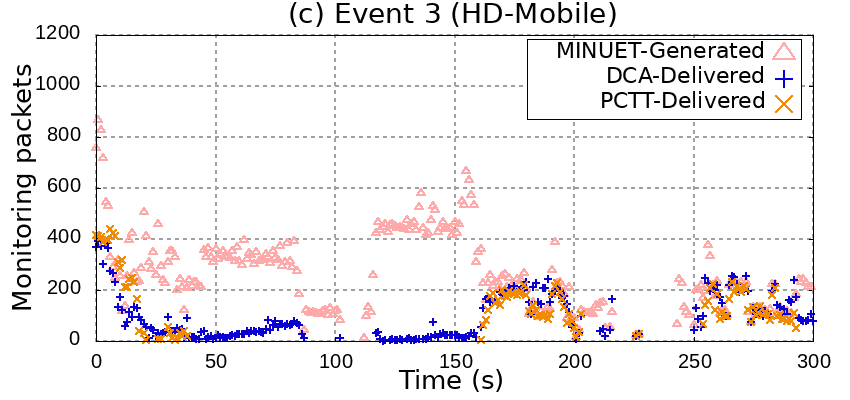}}\\
\subfloat{\includegraphics[draft=false,width=0.76\linewidth]{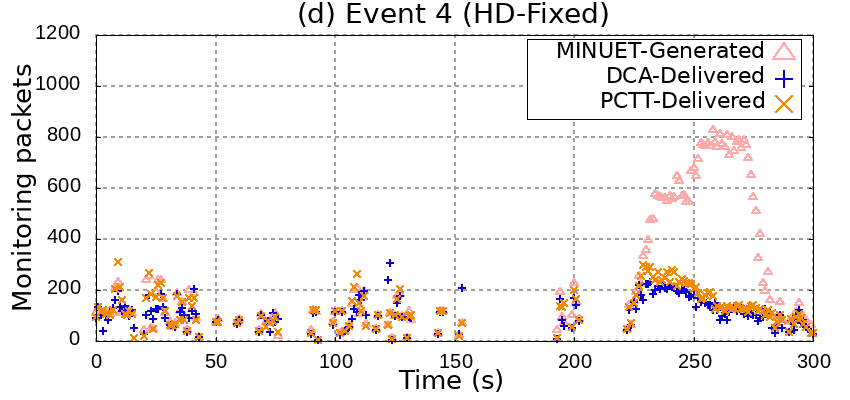}}\\
\caption{The number of generated and delivered monitoring packets (Eq. \ref{eq:num_msgs}(b) and \ref{eq:num_msgs}(c)).}
\label{fig:npe}
\end{figure}

Regarding the monitoring activity, Fig.~\ref{fig:npe} depicts the system's capacity to monitor events and deliver monitoring packets. The results do not show a significant difference in the number of delivered packets when using both clustering algorithms, the exception being an intermittent PCTT behavior during the delivery of \textit{EV3} monitoring packets, from 40 to 160 seconds in Fig.~\ref{fig:npe}(c).
Despite that, the results for mobile events \textit{EV1} and \textit{EV3} showed a more continuous behavior over time (Fig.~\ref{fig:npe}(a) and Fig~\ref{fig:npe}(c)), when compared to the results of fixed events \textit{EV2} and \textit{EV4}, in which longer intervals without delivery occur (Fig.~\ref{fig:npe}(b) and Fig~\ref{fig:npe}(d)). This is expected due to the higher number of the same vehicles continuously detecting and cooperating in the network (Fig.~\ref{fig:nvd} and Fig.~\ref{fig:nvc}).
The system was able to deliver packets during approximately 77\% of the time for mobile events \textit{EV1} and \textit{EV3} with DCA; more than 50\% of the time for fixed events \textit{EV2} and \textit{EV4} with both DCA and PCTT; and 39\% of the time with PCTT for fixed event \textit{EV3}. 
Although mobile events presented a similar delivery period in both density scenarios, a larger number of packages is perceived in the LD scenario, due to the higher number of vehicles detecting (Fig.~\ref{fig:nvd}(a)) and collaborating (Fig.~\ref{fig:nvc}(a)). 
So, it is clear that continuous detection, as well as a high number of detecting and cooperating vehicles, have a strong impact on the delivery of monitoring packages to the corresponding authorities.


Redundancy in the process of disseminating and delivering data aims the authorities to receive more accurate information about events. This way, the redundancy level in the delivery of monitoring packages was also measured, i.e. repeated packets delivered at the base stations. We have used Eq. 4(a) and 4(b) to express the percentage of redundant and non-redundant monitoring packets, and Table~\ref{tab:razao_redun_dca_pctt} shows the obtained results for each event, according to the applied clustering algorithm. A higher redundancy rate is achieved by using DCA for mobile and fixed events, as during the phases of announcement and clustering, it provides the formation of larger groups in the system. However, considering the similar delivery rates and the low redundancy rate of the PCTT, it is able to achieve a lower number of packet losses since less overhead is incurred.

\begin{table}[h]
\centering
\footnotesize
\onehalfspacing
\caption{Redundant and non-redundant delivered data (Eq. \ref{eq:redundancy}(a) and \ref{eq:redundancy}(b)).}
\label{tab:razao_redun_dca_pctt}
\begin{tabular}{|c||C{1.5cm}|R{1.5cm}|C{1.5cm}|R{1.5cm}|} 
\hline
\multirow{2}{*}{\begin{tabular}[c]{@{}c@{}}\textbf{Low-density} \\ \textbf{Scenario} \end{tabular}} 
& \multicolumn{2}{c|}{\textbf{EV1}} & \multicolumn{2}{c|}{\textbf{EV2}} \\
& \multicolumn{2}{c|}{$MP_g(\Delta t) = 97528$} & \multicolumn{2}{c|}{$MP_g(\Delta t) = 40080$ } 
\\ 
\hline
 & \textbf{$\mathcal{S}(\Delta t)$}  & \textbf{$\mathcal{R}(\Delta t)$(\%)} & \textbf{$\mathcal{S}(\Delta t)$}  & \textbf{$\mathcal{R}(\Delta t)$(\%)} 
\\ 
\textbf{DCA} & 40850 & 26.10 & 29396 & 20.62
\\
\textbf{PCTT} & 41304 & 1.74 & 29703 & 7.93
\\
\hline
\hline
\multirow{2}{*}{\begin{tabular}[c]{@{}c@{}}\textbf{High-density} \\ \textbf{Scenario} \end{tabular}}
& \multicolumn{2}{c|}{\textbf{EV3}} & \multicolumn{2}{c|}{\textbf{EV4}}
\\
& \multicolumn{2}{c|}{$MP_g(\Delta t) = 70493$ } & \multicolumn{2}{c|}{$MP_g(\Delta t) = 45437$ } \\
\hline
 & \textbf{$\mathcal{S}(\Delta t)$}  & \textbf{$\mathcal{R}(\Delta t)$(\%)} & \textbf{$\mathcal{S}(\Delta t)$} & \textbf{$\mathcal{R}(\Delta t)$(\%)}
\\ 
\textbf{DCA} & 22194 & 31.04 & 19593 & 17.64
\\
\textbf{PCTT} & 17450 & 13.62 & 22268 & 7.50
\\
\hline
\end{tabular}
\end{table}

\begin{figure}[!htb]
\centering
\subfloat{\includegraphics[draft=false,width=0.4\columnwidth]{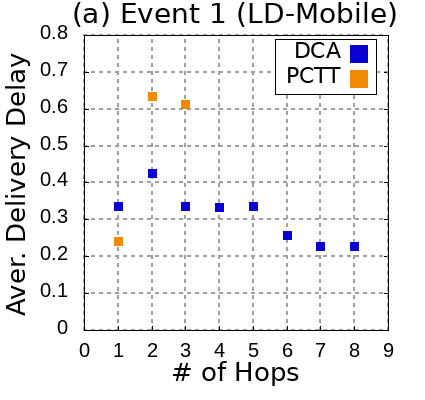}}
\subfloat{\includegraphics[draft=false,width=0.4\columnwidth]{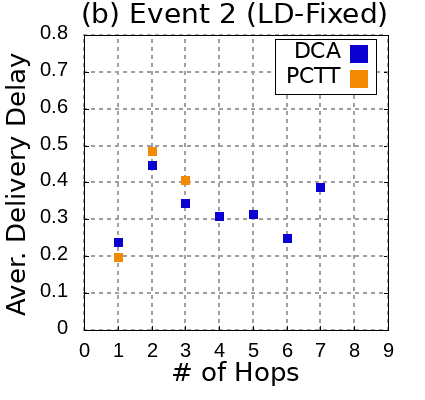}} \\
\subfloat{\includegraphics[draft=false,width=0.4\columnwidth]{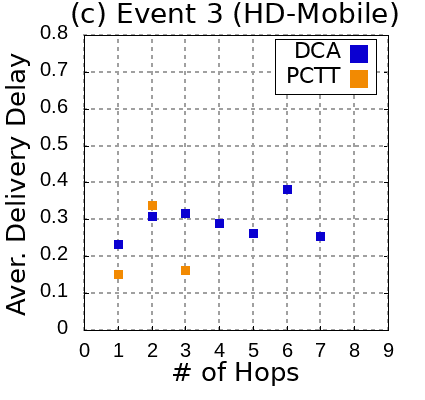}}
\subfloat{\includegraphics[draft=false,width=0.4\columnwidth]{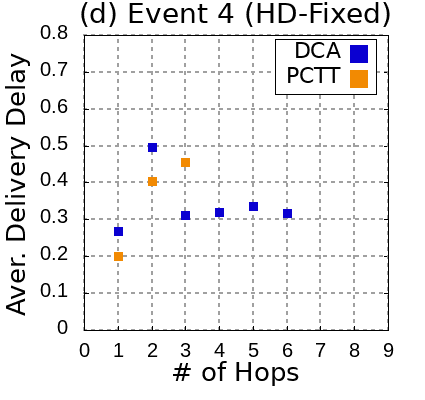}}
\caption{Average delay of the monitoring packets (Eq. \ref{eq:avg_delay}).}
\label{fig:ame}
\end{figure}

Low delays in delivering event information to the corresponding authority imply in making correct and efficient decisions. Hence, Fig.~\ref{fig:ame} presents the average delay of delivered packets (Eq. \ref{eq:avg_delay}), segmented by the number of hops provided by each clustering algorithm. As indicated by Fig.~\ref{fig:ame}(a) and Fig.~\ref{fig:ame}(c), mobile events present a higher number of hops since they form more and larger groups (mainly when using DCA). The average delay remains close to 0.3s in most of the cases, which is a reasonable value for one-way video streaming applications \cite{Szigeti2013}. Also, average delays are usually lower with DCA, despite higher rates of redundant packets and packet loss compared to PCTT. 
The results show that the delay times are affected more by the type of the clustering technique than by the event mobility.

\section{Conclusions} 
\label{sec:conclusions}

The article presented an assessment on how mobility and simultaneity of critical events in urban environments impact the VANET-supported delivery of data stream to a corresponding authority. The monitoring, dissemination, and delivery of information services were analyzed taking into consideration the distribution of the events in the physical space of the environment over time and under different contexts of urban traffic. The MINUET system was used as a clustering-based approach to deliver events' information, with which simulated experiments were conducted. 
Obtained results showed that vehicular traffic holding the same orientation as the mobile event favors a more continuous monitoring of the event.
In addition, results indicate that there is no direct relationship between traffic density and the number of vehicles cooperating in a real traffic environment.
Lastly, the MINUET system was able to cooperatively disseminate events' data considering their mobility and coexistence, and under different vehicular traffic densities.
These results are expected to encourage the development of new data solutions that take into account the impact of moving and simultaneous events have on the dissemination of real-time data in VANETs.



\bibliographystyle{spphys}       
\bibliography{references.bib}   

\begin{thebibliography}{10}
\providecommand{\url}[1]{{#1}}
\providecommand{\urlprefix}{URL }
\expandafter\ifx\csname urlstyle\endcsname\relax
  \providecommand{\doi}[1]{DOI \discretionary{}{}{}#1}\else
  \providecommand{\doi}{DOI \discretionary{}{}{}\begingroup
  \urlstyle{rm}\Url}\fi

\bibitem{Zhang2016}
N.~Zhang, H.~Huang, B.~Su, Comprehensive analysis of information dissemination
  in disasters, Physica A: Statistical Mechanics and its Applications
  \textbf{462}, 846  (2016)

\bibitem{Costa2020}
D.G. Costa, F.P. {de Oliveira}, A prioritization approach for optimization of
  multiple concurrent sensing applications in smart cities, Future Generation
  Computer Systems \textbf{108}, 228  (2020)

\bibitem{TangY2019}
Y.~Tang, S.~Huang, Assessing seismic vulnerability of urban road networks by a
  bayesian network approach, Transportation Research Part D: Transport and
  Environment \textbf{77}, 390  (2019)

\bibitem{Cantillo2019}
V.~Cantillo, L.F. Macea, M.~Jaller, Assessing vulnerability of transportation
  networks for disaster response operations, Networks and Spatial Economics
  \textbf{19}(1), 243 (2019)

\bibitem{TangP2019}
P.~Tang, Q.~Xia, Y.~Wang, Addressing cascading effects of earthquakes in urban
  areas from network perspective to improve disaster mitigation, International
  Journal of Disaster Risk Reduction \textbf{35}, 101065 (2019)

\bibitem{Amiri2020}
I.S. Amiri, J.~Prakash, M.~Balasaraswathi, V.~Sivasankaran, T.V.P.
  Sundararajan, M.H.D.N. Hindia, V.~Tilwari, K.~Dimyati, O.~Henry, Dabpr: a
  large-scale internet of things-based data aggregation back pressure routing
  for disaster management, Wireless Networks \textbf{26}(4), 2353 (2020)

\bibitem{Ferranti2019}
L.~{Ferranti}, S.~{D'Oro}, L.~{Bonati}, E.~{Demirors}, F.~{Cuomo},
  T.~{Melodia}, Hiro-net: Self-organized robotic mesh networking for internet
  sharing in disaster scenarios, in \emph{2019 IEEE 20th International
  Symposium on "A World of Wireless, Mobile and Multimedia Networks" (WoWMoM)}
  (2019), pp. 1--9

\bibitem{Immich2019}
R.~Immich, E.~Cerqueira, M.~Curado, Efficient high-resolution video delivery
  over vanets, Wireless Networks \textbf{25}(5), 2587 (2019)

\bibitem{Hilal2020}
N.~{Hilal}, A.~{Yurdakul}, Model-based design of a roadside unit for emergency
  and disaster management, in \emph{NOMS 2020 - 2020 IEEE/IFIP Network
  Operations and Management Symposium} (2020), pp. 1--6

\bibitem{Chou2017}
Y.H. Chou, T.H. Chu, S.Y. Kuo, C.Y. Chen, An adaptive emergency broadcast
  strategy for vehicular ad hoc networks, IEEE Sensors Journal pp. 1--1 (2017)

\bibitem{Shah2019}
S.A. {Shah}, et~al., Towards disaster resilient smart cities: Can internet of
  things and big data analytics be the game changers?, IEEE Access \textbf{7},
  91885 (2019)

\bibitem{Astarita2020}
V.~Astarita, V.P. Giofrè, G.~Guido, G.~Stefano, A.~Vitale, Mobile computing
  for disaster emergency management: Empirical requirements analysis for a
  cooperative crowdsourced system for emergency management operation, Smart
  Cities \textbf{3}(1), 31 (2020)

\bibitem{Boukerche2019}
A.~{Boukerche}, Smart disaster management and responses for smart cities: A new
  challenge for the next generation of distributed simulation systems, in
  \emph{2019 IEEE/ACM 23rd Int. Symposium on Distributed Simulation and Real
  Time Applications (DS-RT)} (2019), pp. 1--2

\bibitem{Shrestha2018}
R.~{Shrestha}, R.~{Bajracharya}, S.Y. {Nam}, Centralized approach for
  trustworthy message dissemination in vanet, in \emph{IEEE/IFIP Network
  Operations and Management Symposium (NOMS)} (2018), pp. 1--5

\bibitem{Li2019}
Z.~Li, Y.~Song, J.~Bi, Cadd: connectivity-aware data dissemination using node
  forwarding capability estimation in partially connected vanets, Wireless
  Networks \textbf{25}(1), 379 (2019)

\bibitem{khakpour_using_2017}
S.~Khakpour, R.W. Pazzi, K.~El-Khatib, Using clustering for target tracking in
  vehicular ad hoc networks, Vehicular Communications \textbf{9}, 83 (2017)

\bibitem{Kim2019}
S.~Kim, Effective crowdsensing and routing algorithms for next generation
  vehicular networks, Wireless Networks \textbf{25}(4), 1815 (2019)

\bibitem{Adam2020}
M.S. Adam, M.H. Anisi, I.~Ali, Object tracking sensor networks in smart cities:
  Taxonomy, architecture, applications, research challenges and future
  directions, Future Generation Computer Systems \textbf{107}, 909  (2020)

\bibitem{Palmieri2016}
F.~Palmieri, M.~Ficco, S.~Pardi, A.~Castiglione, A cloud-based architecture for
  emergency management and first responders localization in smart city
  environments, Computers \& Electrical Engineering \textbf{56}, 810  (2016)

\bibitem{Yelure2020}
B.~Yelure, S.~Sonavane, Performance of routing protocols using mobility models
  in vanet, in \emph{Next Generation Information Processing System}, ed. by
  P.~Deshpande, A.~Abraham, B.~Iyer, K.~Ma (Springer Singapore, Singapore,
  2020), pp. 272--280

\bibitem{Harri2009}
J.~{Harri}, F.~{Filali}, C.~{Bonnet}, Mobility models for vehicular ad hoc
  networks: a survey and taxonomy, IEEE Communications Surveys Tutorials
  \textbf{11}(4), 19 (2009)

\bibitem{Medeiros2019}
D.S.V. Medeiros, D.A.B. Hernandez, M.E.M. Campista, et~al, Impact of relative
  speed on node vicinity dynamics in vanets, Wireless Networks \textbf{25}(4),
  1895 (2019)

\bibitem{Goyal2019}
A.K. Goyal, G.~Agarwal, A.K. Tripathi, Network architectures, challenges,
  security attacks, research domains and research methodologies in vanet: A
  survey, International Journal of Computer Network and Information Security
  \textbf{10}(10), 37 (2019)

\bibitem{Ramakrishnan2017}
B.~Ramakrishnan, R.~Bhagavath~Nishanth, M.~Milton~Joe, M.~Selvi, Cluster based
  emergency message broadcasting technique for vehicular ad hoc network,
  Wireless Networks \textbf{23}(1), 233 (2017)

\bibitem{Balakumar2019}
C.~Balakumar, E.~Karthikeyan, A review analysis on emergency data dissemination
  techniques in vehicular adhoc networks, Int. Journal of Scientific \&
  Technology Research \textbf{8}, 1209 (2019)

\bibitem{Dar2010}
K.~Dar, M.~Bakhouya, J.~Gaber, M.~Wack, {Evaluating information dissemination
  approaches in VANETs}, in \emph{Proc. 7th ACM International Conference on
  Pervasive Services (ICPS 2010)} (2010), pp. 120--125

\bibitem{Suthaputchakun2015}
C.~{Suthaputchakun}, Z.~{Sun}, M.~{Dianati}, Impact of propagation environments
  on emergency message dissemination in vanets, in \emph{7th Int. Conf. on
  Ubiquitous and Future Networks} (2015), pp. 361--366

\bibitem{Paranjothi2020}
A.~Paranjothi.
\newblock {Performance Analysis of Message Dissemination Techniques in VANET
  using Fog Computing} (2020).
\newblock \urlprefix\url{https://arxiv.org/abs/2003.04354}

\bibitem{Andrade2020}
E.~Andrade, K.~Veloso, N.~Vasconcelos, A.~Santos, F.~Matos, Cooperative
  monitoring and dissemination of urban events supported by dynamic clustering
  of vehicles, Pervasive and Mobile Computing p. 101244 (2020)

\bibitem{benkerdagh_cluster_2019}
S.~Benkerdagh, C.~Duvallet, Cluster-based emergency message dissemination
  strategy for vanet using v2v communication, International Journal of
  Communication Systems \textbf{32}(5), e3897 (2019)

\bibitem{siddiqua_icafe_2019}
A.~Siddiqua, M.A. Shah, H.A. Khattak, I.U. Din, M.~Guizani, icafe: Intelligent
  congestion avoidance and fast emergency services, Future Generation Computer
  Systems \textbf{99}, 365  (2019)

\bibitem{Derder2019}
A.~Derder, S.~Moussaoui, Z.~Doukha, A.~Boualouache, An online target tracking
  protocol for vehicular ad hoc networks, Peer-to-Peer Networking and
  Applications \textbf{12}, 969  (2019)

\bibitem{Tal2016}
I.~Tal, P.~Kelly, G.M. Muntean, A novel direction-based clustering algorithm
  for vanets, in \emph{2016 23rd Int. Conference on Telecommunications (ICT)}
  (2016), pp. 1--5

\bibitem{Szigeti2013}
T.~Szigeti, et~al., \emph{{End-to-end Qos Network Design}: Quality of Service
  for Rich-Media \& Cloud Networks} (Cisco Press, Indianapolis, IN, 2013)

\end{thebibliography}

%
%

\end{document}